\documentclass[11pt, a4paper]{article}
\usepackage{graphicx}
\usepackage{amssymb}
\usepackage{amsmath}
\usepackage{slashed}
\usepackage{multirow}
\usepackage[backref=none]{ hyperref}
\usepackage{color}
\usepackage{cite}
\usepackage{csquotes}

\usepackage[utf8x]{inputenc}

\usepackage{authblk}

\newcommand{\be}{\begin{equation}}
\newcommand{\ee}{\end{equation}}

\newcommand{\bea}{\begin{align}}
\newcommand{\eea}{\end{align}}
\newcommand{\beq}{\begin{equation}}
\newcommand{\eeq}{\end{equation}}
\newcommand{\nbea}{\begin{align*}}
\newcommand{\neea}{\end{align*}}
\newcommand{\nbeq}{\begin{equation*}}
\newcommand{\neeq}{\end{equation*}}
\newcommand{\bear}{\begin{eqnarray}}  
\newcommand{\eear}{\end{eqnarray}}  

 \definecolor{cadmiumgreen}{rgb}{0.0, 0.42, 0.24}

\definecolor{darkred}{rgb}{0.7,0.,0.3}

\newcommand{\AddrUnina}{\small Dipartimento di Fisica ``E. Pancini", 
Universit\`a di Napoli Federico II \\ and INFN, Sezione di Napoli, via Cinthia, 80126 Napoli (NA), Italy 
}

\title{ 
{ \bf Off-shell probes of the Higgs Yukawa couplings: light quarks and charm}}
\author{Natascia Vignaroli}
\affil{\AddrUnina}
\date{}

\begin{document}
\maketitle

\begin{abstract}{\normalsize 
We review the present status and the future prospects for the measurements of the Higgs Yukawa couplings to light quarks and charm.
A special focus is given to new proposed off-shell probes, which offer promising and complementary opportunities to test light quark Yukawas in triboson final states. Additionally, a new off-shell strategy to test the charm Yukawa coupling in the final state with two bosons plus a charm and a jet is considered. First estimates for the HL-LHC and the FCC-hh sensitivities on the channel are presented, showing encouraging results. 
 }
\end{abstract}

\newpage

\begin{section}{The Higgs Yukawa couplings}
The discovery of the Higgs boson at CERN ten years ago \cite{ATLAS:2012yve, CMS:2012qbp} marked a breakthrough in our understanding of the physics of fundamental interactions, confirming the existence of the Brout-Englert-Higgs mechanism for the electroweak symmetry breaking (EWSB). However, it leaves us with unresolved questions in the Standard Model (SM), above all the non-explanation of the origin of the EWSB, the precarious nature of the associated potential, the unknown origin of the neutrino masses, the Higgs naturalness problem and the mysterious pattern of Yukawa couplings. The investigation of the properties of the Higgs, including its interactions, and in general of the EWSB sector is the primary objective of the LHC, which is approaching the upgrade to its high luminosity phase (HL-LHC) \cite{Cepeda:2019klc}, as well as that of future planned experiments at the energy frontiers, e.g. the FCC \cite{FCC:2018byv, FCC:2018evy, FCC:2018vvp}, the ILC \cite{Behnke:2013xla} or a multi-TeV muon collider \cite{InternationalMuonCollider:2022qki}.
ATLAS \cite{ATLAS:2021vrm, ATLAS:2021qwz, ATLAS:2020jwz, ATLAS:2017ztq} (Fig. \ref{fig:Yuk-H}) and CMS \cite{ CMS:2020cga, CMS:2019pyn, CMS:2018nsn} measurements so far have given firm evidence of Higgs interactions to top, bottom and tau. The corresponding Yukawa couplings are measured with an accuracy of the order of 10\% \cite{ATLAS:2021vrm}. There has been recently the first evidence for Higgs decay to muons \cite{ATLAS:2020fzp, CMS:2020xwi}, which represents the first probe of the Higgs interaction with the second generation of fermions and could give hints on scenarios beyond the SM (BSM) \cite{Vignaroli:2009vt}. Crucial indication on the Higgs role in the mass generation of 1$^{\rm st}$ and 2$^{\rm nd}$ families
would come from the challenging measurement of the light quark Yukawas. These are extremely difficult measurements but several techniques, which we are going to review, have been recently developed to improve the detectability of the Higgs couplings to light quarks and to set bounds on possible deviations induced by physics beyond the SM. The current best constraints are placed by considering global fits to the Higgs strength, which is modified by shifts on the quark Yukawas ($\delta y_q$):
\begin{equation}
\mu=\frac{1}{1+\sum_q (2 \delta y_q + \delta y^2_q ) {\rm Br} (h \to qq)_{\rm SM}}
\end{equation}
where we are neglecting modifications to the Higgs production rate, since this would be affected only by $\delta y_q \gtrsim \mathcal{O}(10^3)$.
Using the most recent measurements from CMS \cite{CMS:2020gsy} and ATLAS \cite{ATLAS:2020qdt}, respectively $\mu=1.06\pm0.07$ and $\mu=1.02^{+0.07}_{-0.06}$, one can set the 95\% C.L. bounds: 
\begin{align}
\begin{split}
 & \delta y_d < 400 \; ,  \delta y_u < 820 \;  , \delta y_s < 19 \quad  {\rm (ATLAS) }  \\
  & \delta y_d < 450 \; , \delta y_u < 930 \; , \delta y_s < 22 \quad  {\rm (CMS) }
\end{split}
\end{align}

\begin{figure}[h!]
\centering
\includegraphics[scale=0.25]{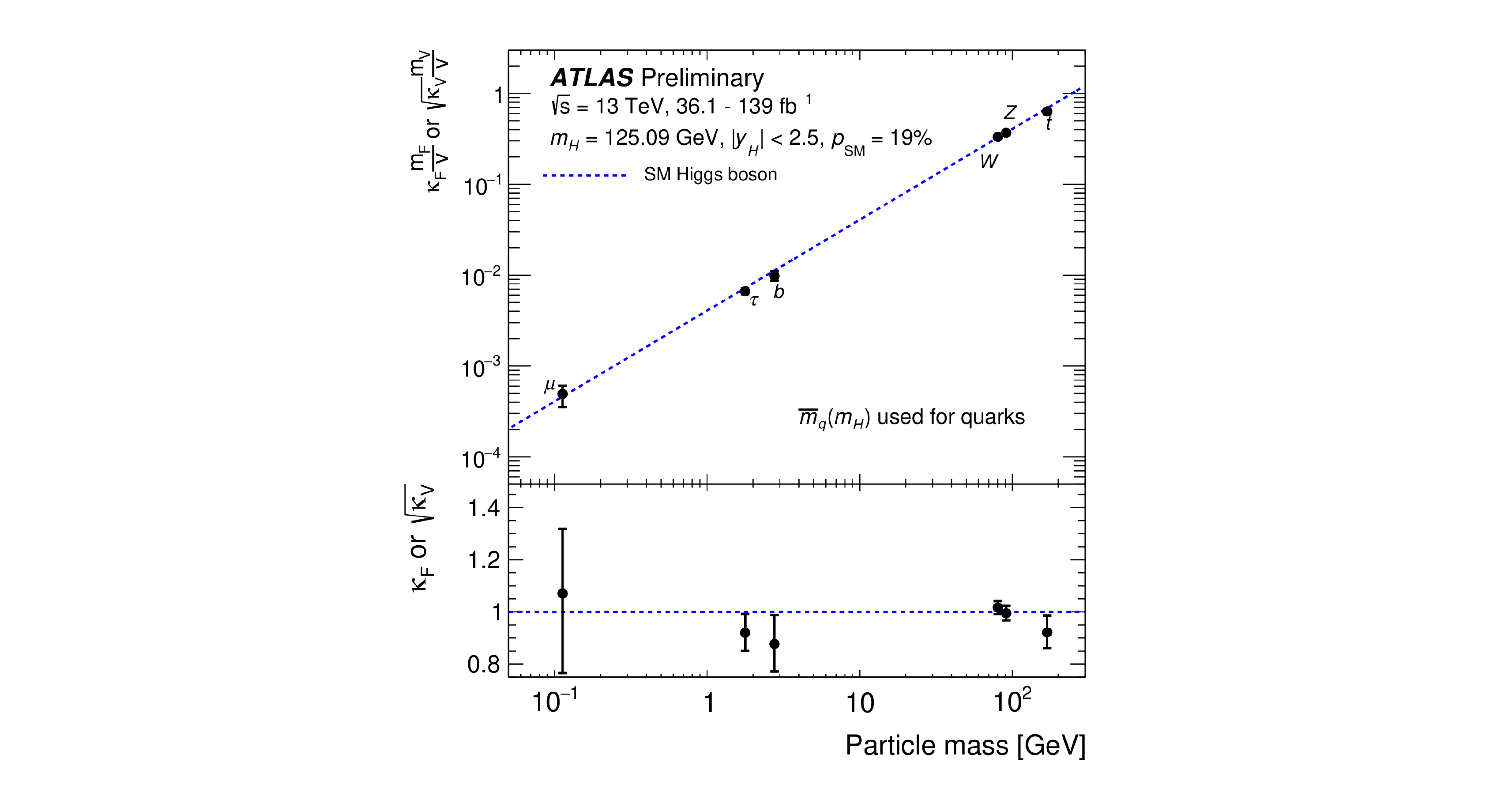} 
\caption{ \small Measurements of Higgs Yukawa couplings confronted with the SM prediction. Plot extracted from the ATLAS study in Ref. \cite{ATLAS:2021vrm}.}
\label{fig:Yuk-H}
\end{figure}

Considering the expectation for the High-Luminosity LHC of measuring the total Higgs signal strength with an error of order 2-3\% \cite{Cepeda:2019klc}, these bounds from global fits could be improved up to

\begin{equation}
 \delta y_d \lesssim 340 \; ,  \delta y_u \lesssim 700 \;  , \delta y_s \lesssim 17 \quad  \text{ (HL-LHC) } \, .
\end{equation}

Note that the different partonic content of the proton has a significant impact on the limits, and the improvement at HL-LHC might depend on the knowledge of the parton distribution functions (PDFs) at that time. Complementary alternative strategies have been proposed to test the light quark Yukawas. These include the study of rare Higgs decays into vector mesons, which gives limits $\delta y_q \lesssim 10^6$  \cite{Kagan:2014ila}, the analysis of the $W^\pm h$ charge asymmetry, with a sensitivity $\delta y_d \lesssim 1300$ \cite{Yu:2016rvv} and, in particular,  the study of the double Higgs production channel \cite{Alasfar:2019pmn} and of the Higgs kinematics ($p_T$ and rapidity) \cite{Soreq:2016rae}, which show competitive sensitivities with those from global fits at the HL-LHC:

\begin{align}
\begin{split}
 & \delta y_d \lesssim 850 \; ,  \delta y_u \lesssim 1200  \quad  \text{(double Higgs prod.) }  \\
  & \delta y_d \lesssim 380 \; , \delta y_u \lesssim 640 \quad \text{ (Higgs kinematics) }
\end{split}
\end{align}

 All of these techniques represent  {\it  on-shell Higgs} probes. We will instead discuss novel strategies which relies on the study of {\it off-shell Higgs} channels, following the idea of ``measuring the Higgs couplings without the Higgs" \cite{Henning:2018kys}.  The key observation behind this approach relies on the fact that modifications to the Higgs couplings affect the delicate cancellations which avoid violation of perturbative unitarity at high energy in scatterings involving electroweak gauge bosons. This leads to measurable energy-growing effects.

\begin{figure}[t]
\centering
\includegraphics[width=0.26\textwidth]{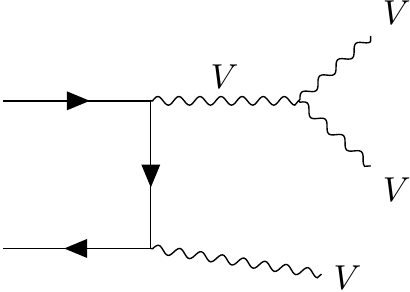} 
\hspace{6mm}
\includegraphics[width=0.26\textwidth]{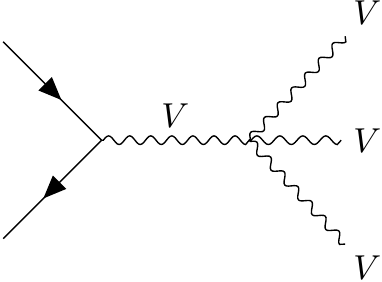} 
\hspace{6mm}
\includegraphics[width=0.26\textwidth]{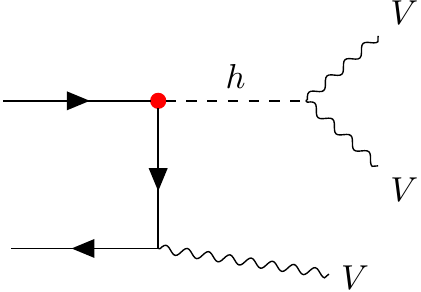} 
\caption{ \small  Feynman diagrams in the unitary gauge contributing to the triple electroweak gauge boson channel, $q \bar q \to VVV$.  \label{fig:vvv_diagrams}
} 
\end{figure}

\section{Light quark Yukawas in triboson final states}

Modifications of light quark Yukawas lead to energy-growing effects in the  {\it off-shell Higgs} triboson channel, which can be then analyzed to put constraints on $\delta y_q$. Fig.~\ref{fig:vvv_diagrams} shows the leading tree-level Feynman diagrams contributing to the triple electroweak gauge boson channel, $q \bar q \to VVV$. In the SM, the Higgs exchange diagrams cancel the bad high-energy growing behavior of the remaining diagrams. 
Modifications of the Higgs Yukawa coupling spoil these delicate cancellations and lead to energy-growing amplitudes. 
We review the main results of the study in \cite{Falkowski:2020znk}, which relies on this effect. 

The framework of the SM effective field theory (SMEFT) is adopted in \cite{Falkowski:2020znk}. In particular, this sub-set of dimension-6 gauge-invariant operators are assumed to encode the dominant BSM contribution to the Higgs Yukawa couplings to light quarks:   %
\begin{equation}
\label{eq:TH_d6}
 {\cal L}_{\rm SMEFT} \supset   - \frac{Y_u |H|^2 }{v^2} \bar u_R Q_{1,L} H  
 - \frac{Y_d |H|^2 }{v^2}  \bar d_R H^\dagger Q_{1,L}
  - \frac{Y_s |H|^2 }{v^2}  \bar s_R H^\dagger Q_{2,L}
+{\rm h.c.} \,  ,  
\end{equation}
where $Q_{1,L} = (u_L,d_L)$ and $Q_{2,L} = (c_L,s_L)$ represent the left-handed $1^{\rm st}$ and $2^{\rm nd}$ generation SM quark doublets, 
$H$ is the Higgs doublet, and $v\approx246$~GeV is the Higgs VEV. 
The parameters  $Y_q$ are assumed to be real, 
for simplicity. An example of a BSM scenario that can generate the operators in \eqref{eq:TH_d6} is that of a new strong dynamics including a pseudo-Nambu-Goldstone composite Higgs \cite{Agashe:2004rs} with decay constant $f$, where Higgs non-linearities can give shifts of the order $v^2/f^2$ to the Yukawa couplings. More in general, the operators  in \eqref{eq:TH_d6} can be generated by heavy vector-like quarks (also typically predicted in composite Higgs models) with masses of order $v/\sqrt{|Y_q|}$, which mix with the SM fermions after the EWSB.

Yukawa couplings are parametrized as 
\begin{equation}
\label{eq:TH_yukawa}
{\cal L} \supset  - \frac{h}{v} \sum_{q=u,d,s} m_q \big ( 1 + \delta y_q \big )  \bar q q .
\end{equation}
Modifications to the Yukawas with respect to the SM values are encoded in the parameters $\delta y_q$, which are related to the coefficients of the operators in Eq.~\eqref{eq:TH_d6} by 
\beq
\label{eq:deltayq_def}
\delta y_q  =   \frac{Y_q}{y_q^{\rm SM}}\, .
\eeq
Here the SM Yukawa couplings are defined as $y_q^{\rm SM} \equiv  \sqrt 2  m_q /v$. \footnote{$y_q^{\rm SM}$ are evaluated at the Higgs mass scale.}      

The leading effect of the modification of the light quark Yukawas
 is manifested in the $q \bar{q} \to 3V_L$ channel. A clear picture of the effects induced by the BSM operators in \eqref{eq:TH_d6} to the triboson 
 process is obtained by adopting a non-unitary gauge description. 
The Higgs doublet can be then parametrized, introducing the would-be-Goldstone bosons $G_i$ and the Higgs field $h$, as 
\begin{equation}
\label{eq:TH_Hdoublet}
H = \frac{1}{\sqrt 2} \begin{pmatrix} i \sqrt 2 G_+ 
\\ v + h + i G_z  \end{pmatrix} \, .
\end{equation}

It is then easy to observe that the effective operators in~\eqref{eq:TH_d6} generate contact interactions between two quarks and three Goldstone bosons:
\begin{eqnarray}
\label{eq:TH_contact}
{\cal L} &  \supset & 
\frac{1}{v^2} \bigg (  G_+ G_- + \frac{1}{2} G_z^2  \bigg ) 
\bigg \{ 
i y_u^{\rm SM }  \delta y_u 
 \left ( \sum_{q' = d,s} \bar u_R  q_L' G_+   - \bar u_R u_L   \frac{G_z}{\sqrt 2} \right )   
\nonumber \\   & +  & 
 i  \sum_{q' = d,s}  y_{q'}^{\rm SM}  \delta y_{q'} 
 \left ( \bar q_R' u_L  G_-  +   \bar q_R'  q_L' \frac{G_z}{\sqrt 2}    \right )   
  + {\rm h.c.} 
  \bigg \} .
 \end{eqnarray}

 These interactions characterize the ${\cal M}(q \bar q \to G G G)$ amplitudes which, by virtue of the equivalence theorem ~\cite{Lee:1977eg}, approximate the high-energy behavior of the  ${\cal M}(q \bar q \to V V V)$ amplitudes. At high energies, $\sqrt{s} \gg m_Z$, the diagram in Fig.~\ref{fig:vvv_goldstonediagram} represents the dominant contribution to the triboson channel.

\begin{figure}[h!]
\centering
\includegraphics[width=0.23\textwidth]{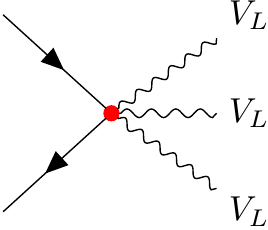} 
\caption{ \small 
The Feynman diagram that gives, in the non-unitary gauge, the dominant contribution to the ${\cal M}(q \bar q \to G G G)$ amplitude at high energies. By virtue of the equivalence theorem  ${\cal M}(q \bar q \to G G G) \approx {\cal M}(q \bar q \to V_L V_L V_L)$ at $\sqrt{s} \gg m_Z$.
\label{fig:vvv_goldstonediagram}
} 
\end{figure}

Simple analytic expressions  can be derived for the cross sections of the $q \bar q \to G G G$ processes induced by the interactions in Eq.~\eqref{eq:TH_contact}:
\begin{eqnarray} 
\label{eq:TXS_qqbGzGpGm}
\sigma (q \bar q \to G_z G_+ G_-)   &= &     ( y_q^{\rm SM }  \delta y_q)^2  I(\hat s) \, , \\ \nonumber 
\sigma (q \bar q \to 3 G_z)   & =  &  
\frac{3}{2}     ( y_q^{\rm SM }  \delta y_q)^2  I(\hat s)  , 
\nonumber \\
\sigma  (u \bar q' \to G_+  G_z G_z)  +  \sigma  (q' \bar u \to G_-  G_z G_z)   & =  &   
\frac{1}{2}     \left [  ( y_u^{\rm SM }  \delta y_u)^2 +  ( y_{q'}^{\rm SM }  \delta y_{q'})^2 \right ]   I( \hat s), 
\nonumber \\
\sigma  (u \bar q' \to G_+  G_+ G_-) + \sigma  (q' \bar u \to G_-  G_- G_+) & =  &   
 2  \left [  ( y_u^{\rm SM }  \delta y_u)^2 +  ( y_{q'}^{\rm SM }  \delta y_{q'})^2 \right ]   I(\hat s), \nonumber \\
  I(\hat s) \equiv  \frac{\hat  s}{6144 \pi^3  v^4} \,  ,
\nonumber
\end{eqnarray} 
where $\sqrt{\hat s}$ is the centre-of-mass energy of the parton-level 
quark-antiquark annihilation and $q = u,d,s$, $q' = d,s$.
By the equivalence theorem,  for $\sqrt{\hat s} \gg m_Z$,  these cross sections are approximately equal to those for
the parton-level triple EW gauge boson production, 
 with the identification $G_\pm \to W^\pm_L$ and $G_z \to Z_L$. 
 It is manifest from \eqref{eq:TXS_qqbGzGpGm} the energy-growing behavior of the cross section for the triboson channel, for $\delta y_q \neq0$. In particular
 \begin{equation}\label{eq:energy}
 \sigma(q \bar q \to V_L V_L V_L) \sim {\cal O}\left(\delta y^2_q \,\frac{\hat s}{v^4}\right ) \, .
 \end{equation}

\begin{figure}[h!]
\centering
\includegraphics[width=0.483\textwidth]{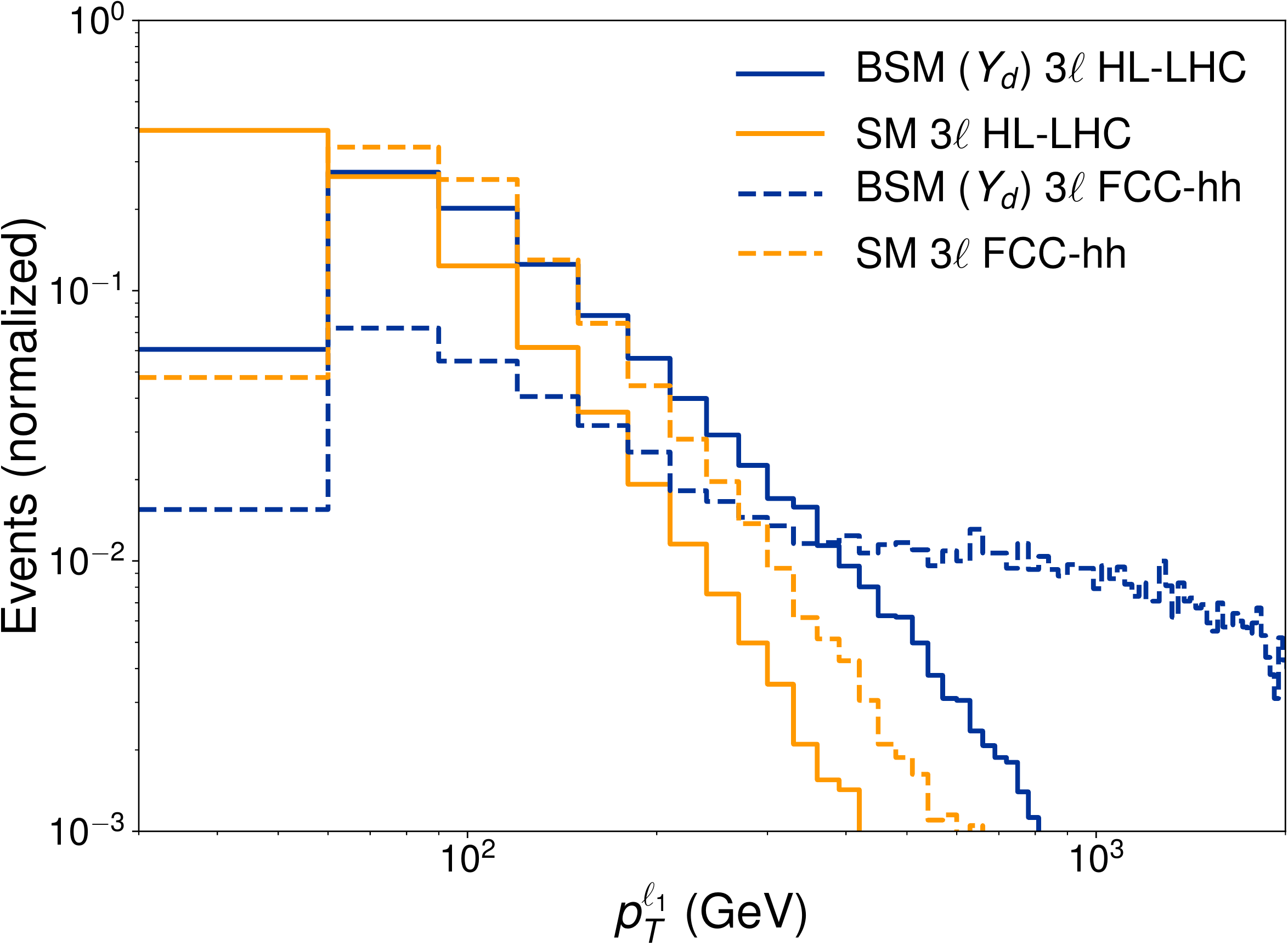} \hspace{1mm}
\includegraphics[width=0.483\textwidth]{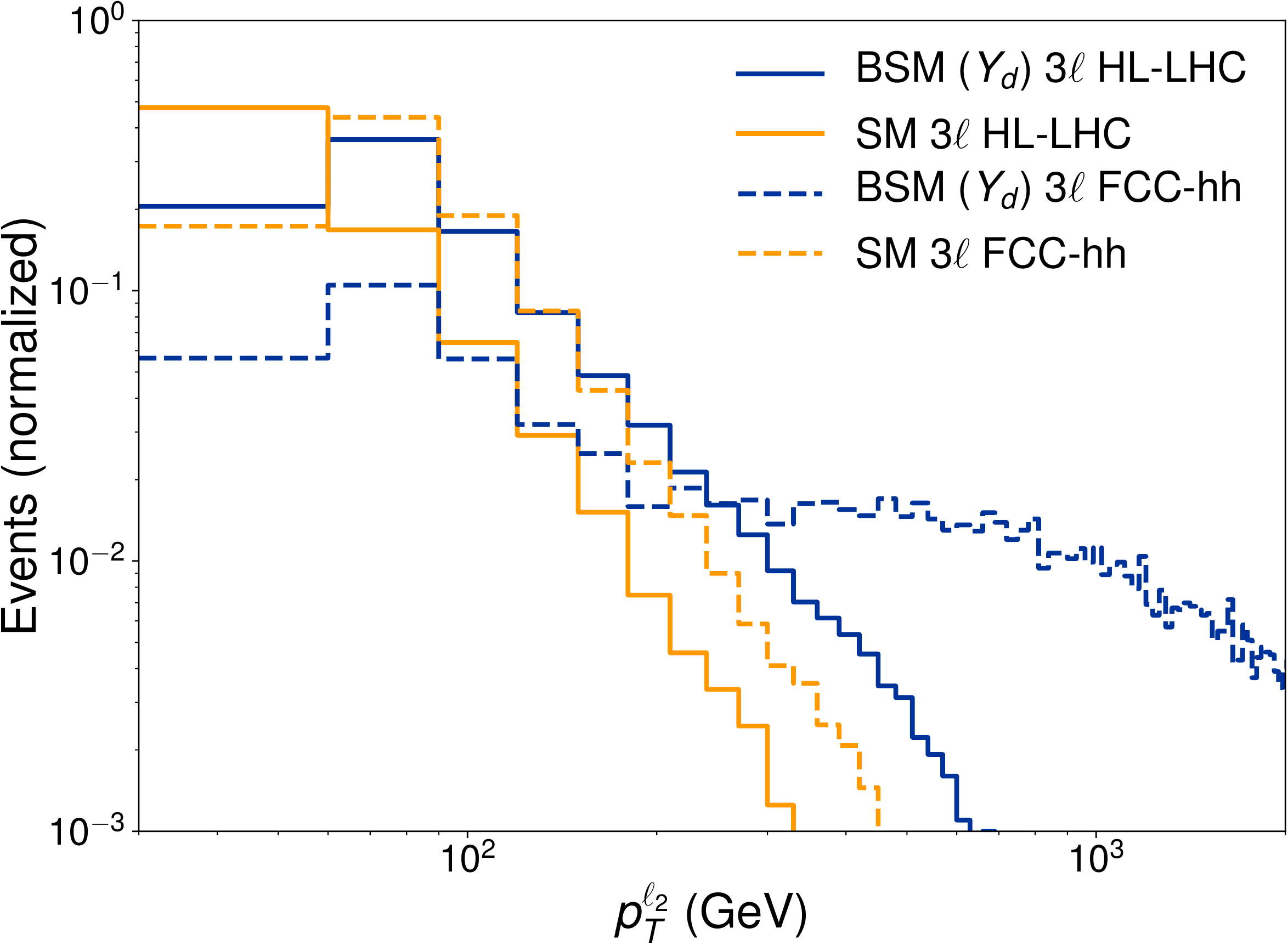} 
\vspace{1mm}

\includegraphics[width=0.483\textwidth]{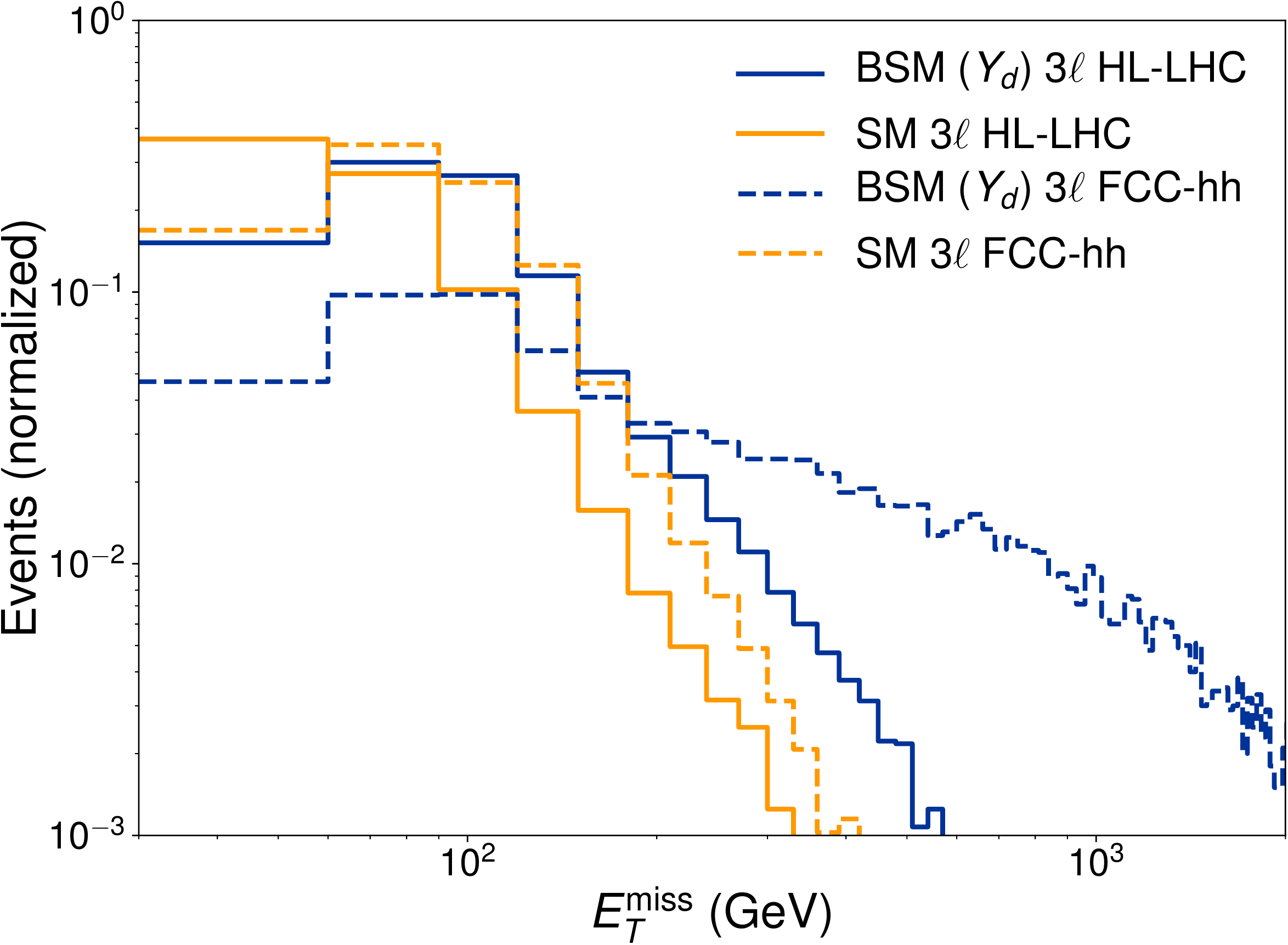} \hspace{2mm}
\includegraphics[width=0.483\textwidth]{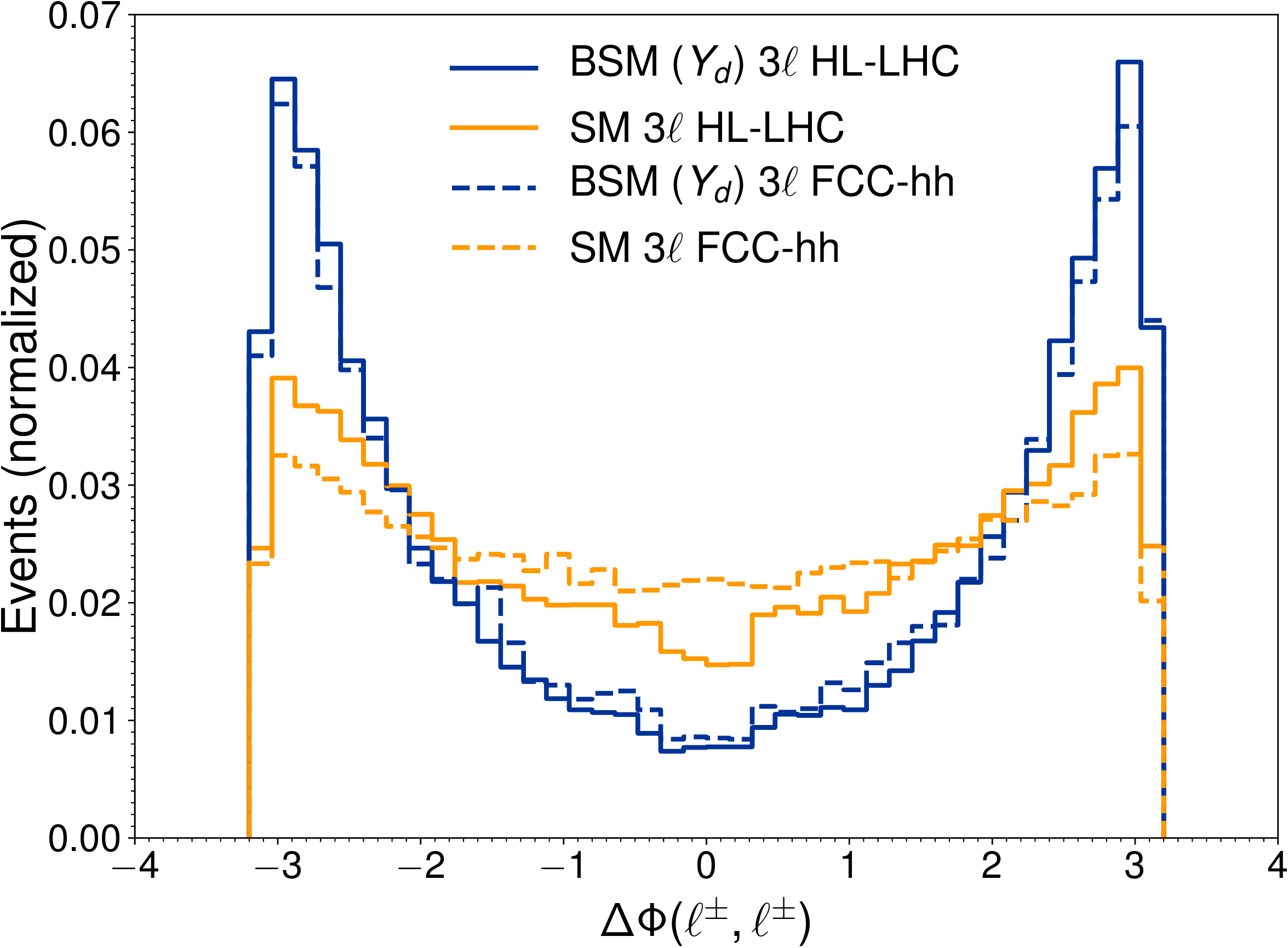}
\caption{ \small 
Differential distributions, normalized to unit area, for the $WWW$ trilepton channel. Top-left: $p_{T}$ of the leading-lepton, top-right: $p_{T}$ of the second-leading lepton, bottom-left: $E^{{\rm miss}}_T$, bottom-right: the azimuthal separation $\Delta \Phi(\ell^{\pm},\ell^{\pm})$. The distributions for the pure BSM triboson signal are indicated by blue lines, those for the SM by yellow lines. Solid lines refer to the 14 TeV LHC, dashed ones to the 100 TeV FCC-hh.
}
\label{fig:3lep-distributions}
\end{figure}

\begin{table*}[t]
\centering
\begin{tabular}{|c|c|c|c|c|}
\hline
{\bf HL-LHC} & SM & BSM ($Y_d=1$)
& BSM ($Y_u=1$) 
& BSM ($Y_s=1$) \\ \hline
$W^+ W^- W^+$ & 152 fb & 3.6 pb & 3.6 pb & 110 fb \\ \hline
$W^+  W^- W^-$ & 87 fb & 1.5 pb & 1.5 pb  & 110 fb \\ \hline
$Z Z W^+$ & 40 fb & 1.0 pb & 1.0 pb & 31 fb \\ \hline
$Z Z W^-$ & 23 fb & 0.43 pb & 0.43 pb & 31 fb \\ \hline
$Z W^+ W^-$ & 191 fb & 1.5 pb & 2.4 pb & 120 fb\\ \hline
$Z Z Z $ & 16 fb & 0.99 pb & 1.7 pb  & 66 fb\\ \hline
\end{tabular}
\end{table*}
\begin{table}[t]
\centering
\hspace{1mm}
\begin{tabular}{|c|c|c|c|c|}
\hline
{\bf FCC-hh} & SM & BSM ($Y_d=1$)
& BSM ($Y_u=1$) 
& BSM ($Y_s=1$) \\ \hline 
$W^+ W^- W^+$ & 2.35 pb & 290 pb & 290 pb & 16 pb\\ \hline
$W^+  W^- W^-$ & 1.76 pb & 140 pb & 140 pb & 16 pb \\ \hline
$Z Z W^+$ & 756 fb & 74 pb & 74 pb & 4.4 pb \\ \hline
$Z Z W^-$ & 579 fb & 36 pb & 36 pb & 4.4 pb \\ \hline
$Z W^+ W^-$ & 3.93 pb & 94 pb & 150 pb & 12 pb \\ \hline
$Z Z Z $ & 231 fb & 110 pb & 180 pb & 11 pb \\ \hline
\end{tabular}
\label{Table_1}
\caption{Cross section values at the $\sqrt{s} = 14$ TeV LHC (upper table), and at the $\sqrt{s} = 100$ TeV FCC-hh (lower table) for different triboson production channels in the SM (computed at NLO in QCD) and for the BSM signals induced by the dimension-6 operators in Eq.~\eqref{eq:TH_d6}, with 
$Y_d = 1$ ($Y_{\neq d} = 0$), $Y_u = 1$ ($Y_{\neq u} = 0$) and $Y_s = 1$ ($Y_{\neq s} = 0$), respectively.
}
\end{table}

Interestingly, the triboson production has been recently measured for the first time at the LHC by CMS \cite{CMS:2020hjs}. This shows the LHC potential to further probe this channel. The main features of the BSM $\delta y_q$ signal are the growing with energy of the cross section, but also a peculiar final state, characterized by the hard emission of gauge bosons and by distinctive angular distributions. The study in  \cite{Falkowski:2020znk} exploits these characteristics to isolate the BSM $\delta y_q$ induced signals from the background. It is clear that this specific search  would significantly benefit from increasing the center of mass energy. It is indeed an ideal case to be studied at the energy frontier experiments, in particular at the FCC-hh.  Tab. \ref{Table_1} shows the cross section values at the HL-LHC and at the FCC-hh of the triboson production channels in the SM and for the BSM signals generated by the operators in Eq.~\eqref{eq:TH_d6} and associated to modifications of down, up and strange Yukawa couplings. Cross sections are computed with {\tt MadGraph5$\_$aMC@NLO}~\cite{Alwall:2014hca}, at the Next-to-Leading Order (NLO) in QCD for the SM~\cite{Dittmaier:2017bnh,Binoth:2008kt} and at LO for the BSM. The BSM signals are generated by using an UFO model \cite{Degrande:2011ua}, which is available at \cite{NV:UFO}.  The study in  \cite{Falkowski:2020znk} performs an analysis of different triboson channels and final states. The most efficient are the $WWW$ channels in same-sign dilepton and trilepton final states. It is however important to consider a combined analysis of several triboson channels to increase the sensitivity and to possibly disentangle the different $\delta y_q$. For example, the study of the neutral $ZZZ$ and $ZW^+ W^-$ channels could in principle distinguish between $\delta y_u$ and $\delta y_d$. The analyses performed in \cite{Falkowski:2020znk} apply a set of cuts on several distinctive observables of the processes. In particular, as shown in Fig. \ref{fig:3lep-distributions}  for the $WWW$ trilepton channel, the $p_T$ of the leptons in the final states, the missing $E_T$, or the angular azimuthal separation between leptons in the final states are  particularly efficient to distinguish the $\delta y_q$ signals, which are characterized by harder final states and a peculiar angular kinematics, from the SM background. The search is then refined by performing a binned-likelihood shape analysis on the $p_T$ distributions of the leading lepton.  More details on the analysis can be found in \cite{Falkowski:2020znk}. Clearly, there is a large room for improvement of this study by considering, for example, combined shape analyses on several interesting $p_T$, $E_T$ and angular distributions. The final results are shown on Table \ref{tab:summary}. The HL-LHC sensitivities on $\delta y_q$ from the analysis of the triboson channels are competitive with those from global fits to Higgs data, with the possibility to test $\delta y_d$ up to order 400. The sensitivities are greatly enhanced at the FCC-hh, which could probe  $\delta y_d$ up to order 30. This corresponds to testing new physics scale for the operators in \eqref{eq:TH_d6} of the order of 10 (3) TeV at the FCC-hh (HL-LHC), as shown in Fig. \ref{fig:scaleplot}.

\begin{figure}[h]
\centering
\includegraphics[width=0.85\textwidth]{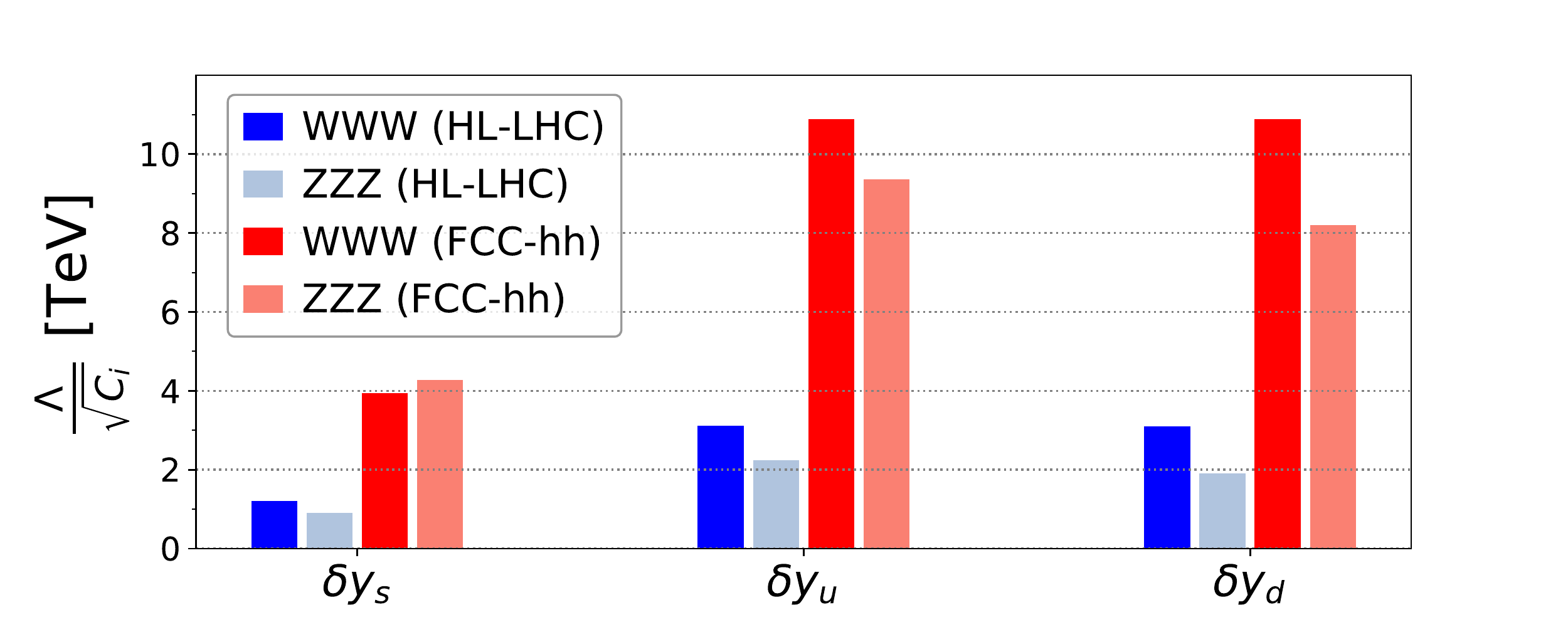} 
\caption{\small  Main results from  Ref. \cite{Falkowski:2020znk}. HL-LHC (blue) and FCC-hh (red) projected 2$\sigma$ reach, for different triboson channels, on the new physics scale $\Lambda$ (with Wilson coefficient $C_i$) at which the dimension-6 Yukawa operators in \eqref{eq:TH_d6} are generated. }
\label{fig:scaleplot}
\end{figure}

\begin{table}[h]
\centering
\begin{tabular}{|c|c|c|c|c|c|c|}
\hline
& \multicolumn{3}{c|}{$WWW$} & \multicolumn{3}{c|}{$ZZZ$} \\
\hline
 & $\ell^\pm \ell^\pm +2\nu +2j$ & $\ell^\pm \ell^\pm \ell^\mp +3\nu$ & Comb. & $4\ell + 2\nu$ & $4\ell + 2j$ & Comb.\\
\hline
$\left. \delta y_d \right .$ & 430 (36) & 840 (54) & 420 (34) & 1500 (65) & 1300 (93) & 1100 (60) \\
\hline
$\left. \delta y_u\right . $ & 850 (71) & 1700 (110) & 830 (68) & 2300 (100) & 1800 (140) & 1600 (92) \\
\hline
$\left. \delta y_s\right .$ & 150 (13) & 230 (33) & 140 (13) & 300 (12) & 290 (16) & 250 (11) \\
\hline
\end{tabular}
\caption{\small HL-LHC (FCC-hh) 2$\sigma$ sensitivities on $\delta y_q$  for the different sub-channels analyzed in  \cite{Falkowski:2020znk}.}
\label{tab:summary}
\end{table}

\section{Off-shell probe of the charm Yukawa}
\label{sec:charm}

\begin{figure}[h!]
\centering
\includegraphics[scale=0.25]{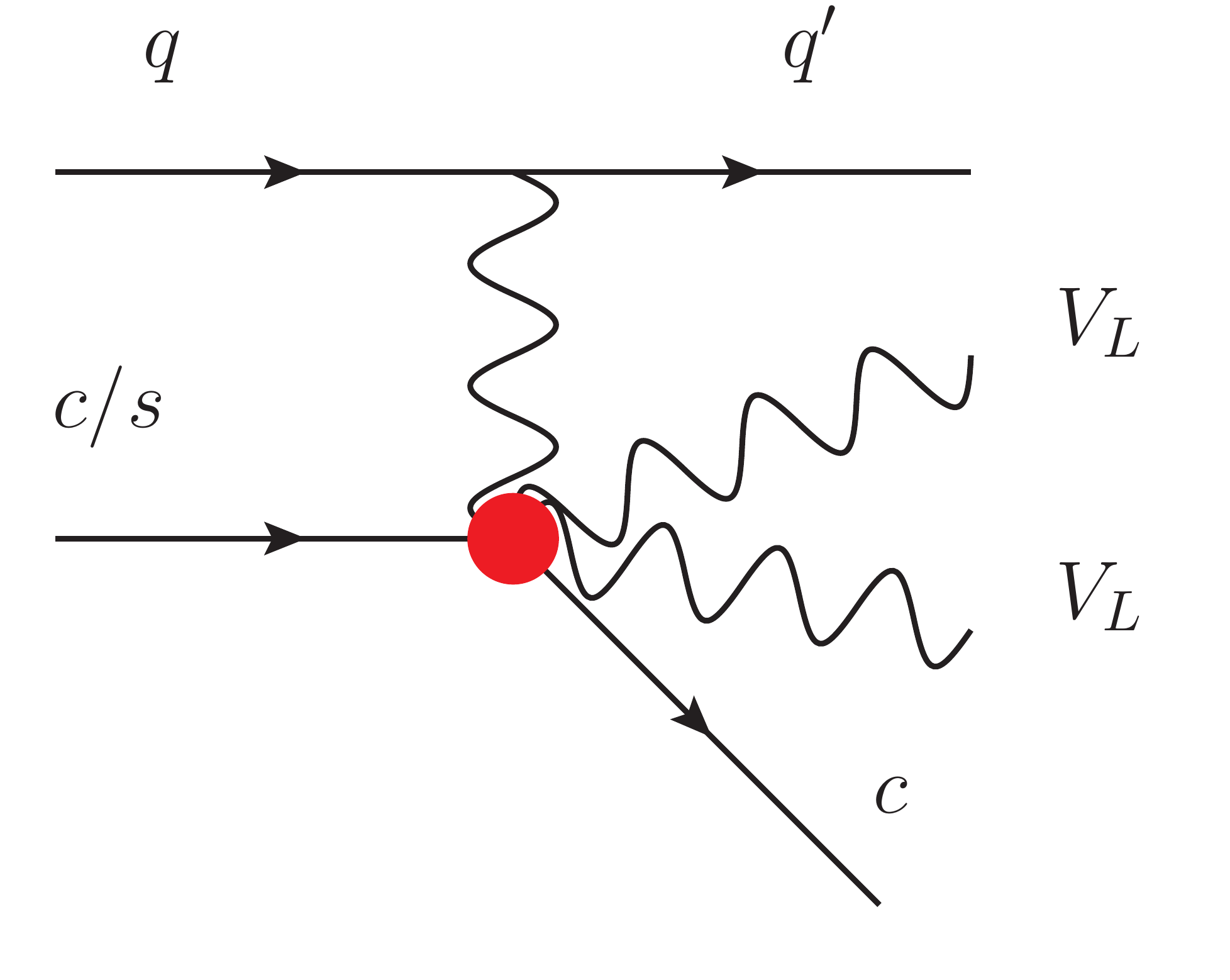} 
\caption{ \small Leading Feynman diagram for the signal of modification to $y_c$ in the non-unitary gauge. The red dot denotes the effective coupling (Eq. (\ref{eq:TH_d6_c})). }
\label{fig:diagram:charm}
\end{figure}

We now focus on probing modifications to the Higgs Yukawa coupling of the charm by analyzing the {\it off-shell Higgs} channel with two gauge bosons plus a tagged charm and a jet in the final state: $VV+c+j$.  The leading Feynman diagram, in the non-unitary gauge, is shown in Fig. \ref{fig:diagram:charm}. In the case of the charm, this two-boson signal is enhanced, compared to the triboson channel considered in the previous section, by the larger parton distribution function, compared to the charm PDF, of the valence quark in the initial state. We will show that this off-shell probe is competitive with other on-shell strategies considered so far: the direct test of the $H \to c\bar{c}$ decay ~\cite{ATLAS:2018tmw, Han:2018juw}, possibly considering the associated Higgs production with a photon \cite{Carlson:2021tes}, the global fit to the Higgs signal strength \cite{Perez:2015aoa}, the indirect test via the radiative process $h\to J/\Psi+\gamma$ \cite{Bodwin:2013gca}, the analysis of the Higgs plus charm channel \cite{Brivio:2015fxa} or the indirect probe from precision Higgs measurements \cite{Coyle:2019hvs}. These latter methods appear to be the most efficient, with the possibility to test, at the 95\% C.L. at the HL-LHC, values: 
\begin{align}
\begin{split}
& |\delta y_c| \lesssim 2.6  \quad  \text{(Higgs plus charm) } \\
& |\delta y_c| \lesssim 2.1  \quad  \text{(Higgs precision measurements) } 
\end{split}
\end{align}

Analogously to the analysis for the triboson channel, we consider modification to the charm Yukawa
\begin{equation}
\delta y_c = \frac{Y_c}{y^{\rm SM}_c} \, 
\end{equation}
induced by the following dim-6 operator

\begin{equation}
\label{eq:TH_d6_c}
 {\cal L}_{\rm SMEFT} \supset  - \frac{Y_c |H|^2 }{v^2} \bar c_R Q_{1,L} H  
+{\rm h.c.} \,  
\end{equation}
We assume $Y_c$ to be a real parameter. As for the case discussed in the previous section, this operator spoils the mechanism of cancellation by the Higgs exchange of the energy-growing diboson amplitudes  and leads to $\left(\delta y^2_c \,\frac{\hat s}{v^4}\right )$ energy enhancement effects, analogous to those described in Eq. \eqref{eq:TXS_qqbGzGpGm}, \eqref{eq:energy}. The BSM contribution can be estimated by considering an expansion of the inclusive cross section in the terms: SM ($Y_c=0$), plus the SM-BSM interference 
term, which is linear in the coefficient $Y_c$, plus the pure BSM term, which is quadratic in $Y_c$:

 \begin{equation}
 \sigma \approx \sigma^{\rm SM}(Y_c=0) + Y_c \, \sigma^{\rm INT}(Y_c=1) + Y^2_c \, \sigma^{\rm BSM}(Y_c=1) \ .
 \end{equation}\\
\noindent
We find the cross section values in the different $VVcj$ channels reported in Tab. \ref{Table_c}. Calculations have been made with  {\tt MadGraph5$\_$aMC@NLO} at LO in QCD by using the UFO model available at \cite{NV:UFO}. Note that, differently from the previous case of the  modifications to the light quark Yukawas, the interference term here is not negligible. This, as we will show, can give an handle to the possibility to test the sign of the modification to the charm Yukawa.

\begin{table*}[h]
\centering
\begin{tabular}{|c|c|c|c|}
\hline
{\bf HL-LHC} & SM ($Y_c=0$) & INT ($Y_c=1$)  & BSM ($Y_c=1$) \\ \hline
$W^+ W^- cj$ & 2.3 pb & 0.58 pb & 63 pb \\ \hline
$W^+ Z cj$ & 0.86 pb & 0.17 pb & 17 pb \\ \hline
$W^- Z cj$ & 0.79 pb & 0.09 pb & 9.1 pb \\ \hline
$Z Z cj$ & 0.19 pb & 0.14 pb & 15 pb \\ \hline
$W^+ W^+ cj$ & 29 fb & 0.42 fb & 94 fb \\ \hline
$W^- W^- cj$ & 23 fb & 0.31 fb & 90 fb \\ \hline
\end{tabular}
\end{table*}
\begin{table}[h]
\centering
\hspace{1mm}
\begin{tabular}{|c|c|c|c|}
\hline
{\bf FCC-hh} & SM ($Y_c=0$) & INT ($Y_c=1$)  & BSM ($Y_c=1$) \\ \hline
$W^+ W^- cj$ & 92 pb & 6.4  pb & 660 pb \\ \hline
$W^+ Z cj$ &  36 pb &  1.8 pb &  190 pb \\ \hline
$W^- Z cj$ & 35 pb &  1.3 pb &  130 pb \\ \hline
$Z Z cj$ & 6.8 pb & 1.6 pb &  180 pb \\ \hline
$W^+ W^+ cj$ & 0.76 pb & 2.8 fb & 3.0 pb \\ \hline
$W^- W^- cj$ & 0.68 pb & 3.2 fb &  3.0 pb \\ \hline
\end{tabular}
\label{Table_c}
\caption{Values of different $VVcj$ cross sections for $\sqrt{s} = 14$ TeV LHC (upper table) and $\sqrt{s} = 100$ TeV FCC-hh (lower table) for the SM and with the addition of the dimension-6 operator from Eq.~\eqref{eq:TH_d6_c}, with 
$Y_c = 1$. We show the cross section for the interference term and the purely BSM quadratic term. Calculations are at LO in QCD and a minimum requirement $p_T\, (j), (c)>$ 20 GeV  is applied.}
\end{table}

\begin{figure}[t]
\centering
\includegraphics[scale=0.42]{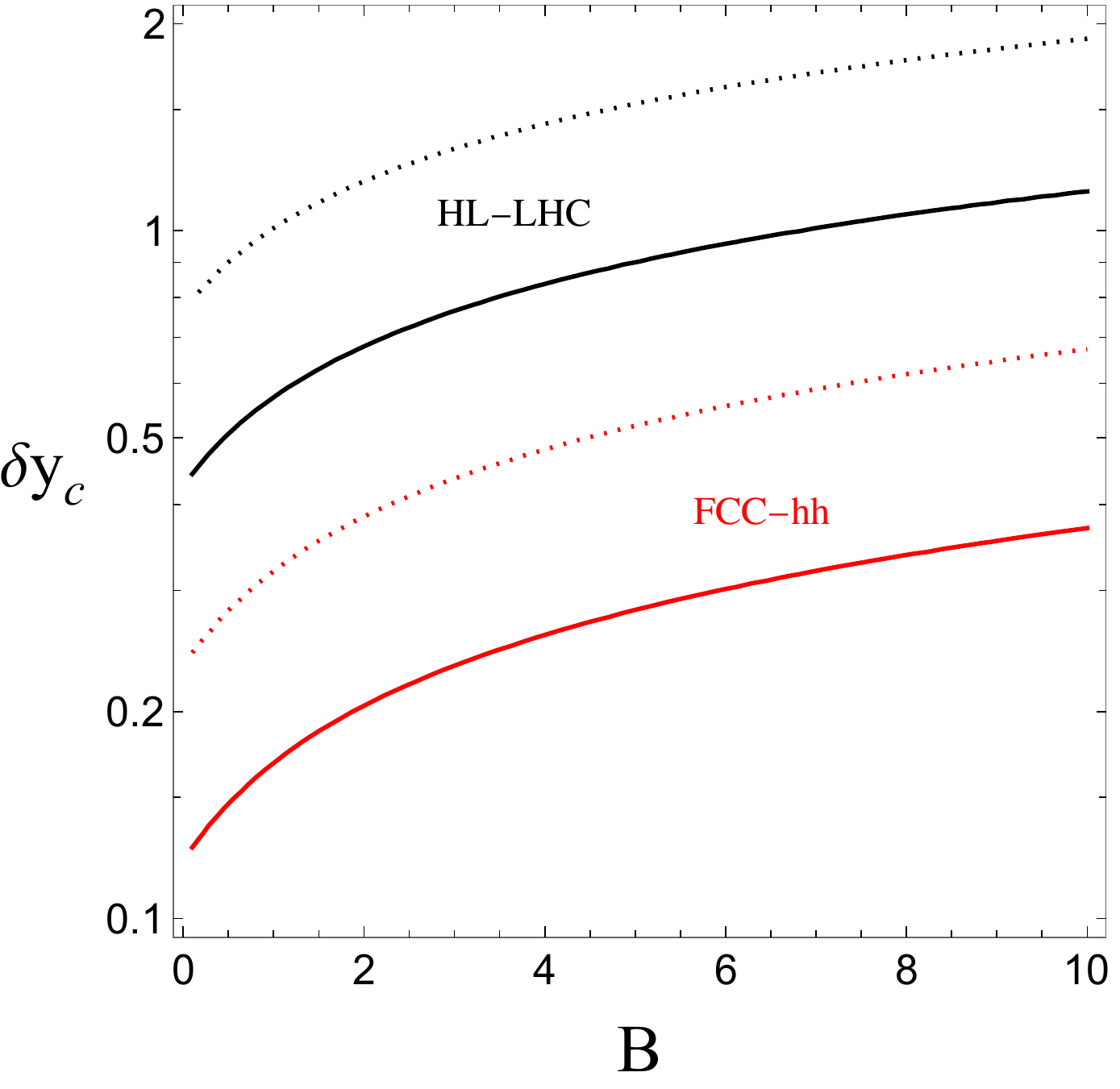}\; \; \includegraphics[scale=0.44]{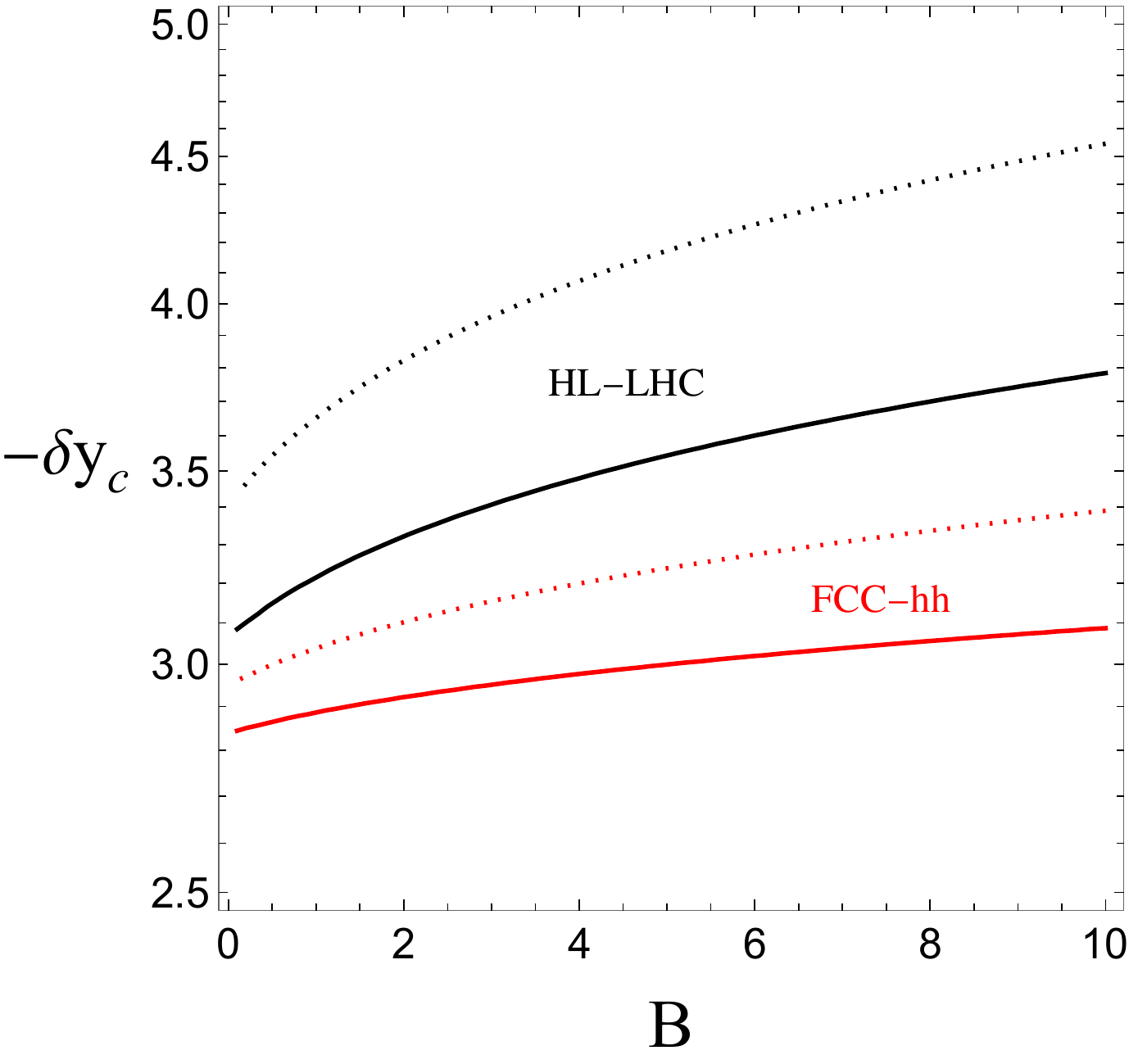} 
\caption{\small  1$\sigma$ (continuous curve) and 2$\sigma$ (dotted curve) HL-LHC (14 TeV, 2$\times$3 ab$^{-1}$) and FCC-hh (100 TeV, 30 ab$^{-1}$) sensitivities on 
$\delta y_c$ in the $VVcj$ channel and semileptonic final state, as functions of an arbitrary amount of reducible background, calculated as a factor $B$ of the irreducible background given by the SM signal (cfr. Eq. (\ref{eq:sigma})).}
\label{fig:Yc}
\end{figure}

At this point, one can derive an estimate of the HL-LHC and FCC-hh sensitivities by considering a naive evaluation of the reducible background, which will contribute to the total background in addition to the irreducible component given by the SM $VVcj$ process.
In particular, we estimate the significance $\sigma$ by applying a background rescaling factor $B$,  which parametrizes the number of reducible background events as a factor of the SM $VVcj $ events: 
\begin{equation}
\sigma = \frac{N_{SM+BSM} - N_{SM}}{\sqrt{ N_{SM}+B \times N_{SM}}} \, .
\label{eq:sigma}
\end{equation}

 By focusing only on the 
semileptonic final state,  $\ell^{\pm}\nu$+$c$+$X$, and considering a c-tagging efficiency of 25$\%$~\cite{ATLAS:2018tmw, Han:2018juw}, we find the 1$\sigma$ and the 2$\sigma$  sensitivities on $\delta y_c$ of the HL-LHC and of the FCC-hh in the $VVcj $ channel shown in
Fig. \ref{fig:Yc}. Sensitivities are presented as functions of the reducible background rescaling factor $B$.
In the hypothesis of a negligible (compared to the SM $VVcj$ process) reducible background, we find the sensitivities on $\delta y_c$ as 

\begin{align}
\begin{split}
&\delta y_c \lesssim \quad 0.43 \,  (1\sigma)\,  -\,0.77 \,  (2\sigma) \qquad \text{(HL-LHC, 2$\times$3 ab$^{-1}$)} \\
&\delta y_c \lesssim \quad 0.12 \,  (1\sigma)\,  -\,0.23 \,  (2\sigma) \qquad \text{(FCC-hh, 30 ab$^{-1}$)}
\end{split}
\end{align}

\begin{align}
\begin{split}
&-\delta y_c \lesssim \quad 3.1 \,  (1\sigma)\,  -\, 3.4 \,  (2\sigma) \qquad \text{(HL-LHC, 2$\times$3 ab$^{-1}$)} \\
&-\delta y_c \lesssim \quad 2.8 \,  (1\sigma)\,  -\, 2.9 \,  (2\sigma) \qquad \text{(FCC-hh, 30 ab$^{-1}$)}
\end{split}
\end{align}

We consider these estimates a useful and encouraging starting point for more refined analysis. Future studies could exploit for example, similarly to what has been discussed for the triboson channels, the peculiar kinematic of the BSM process to better isolate its contribution from the background and minimize in particular the reducible component of the background.
Our results indicate sensitivities on $\delta y_c$ in the diboson channel which could be realistically below order 1 at the HL-LHC, thus competitive and complementary to other charm Yukawa probes, and of the order of 20$\%$ at the FCC-hh. Interestingly, there is also the possibility to test the sign of $\delta y_c$, since, because of interference effects, significances are higher in the case of a positive shift to the charm Yukawa. Negative shifts of the order of 3$y^{\rm SM}_c$ can be probed in the $VVcj$ channel.\\

\end{section}

\begin{section}{Conclusions }

The investigation of the properties of the Higgs boson, and in particular of its interactions and Yukawa couplings to SM fermions is of primary importance for the understanding of high energy physics. Crucial indication on the Higgs role in the mass generation of 1$^{\rm st}$ and 2$^{\rm nd}$ families
would come from the measurement of the light quark Yukawas. We have reviewed a novel technique to improve this challenging test based on the study of the triboson channel, where the Higgs is off-shell. In this approach, the Yukawa couplings are determined indirectly,  by their contributions via virtual Higgs exchange to the triboson process. This method relies on the fact that modifications of the Higgs Yukawas disturb the structure
of the SM and leads to the violation of perturbative unitarity at high energy. 
The study of the triboson channel gives results competitive and complementary to other on-shell Higgs probe. In particular, the HL-LHC could test modification to the down (up) quark, $\delta y_d\, (\delta y_u)$, up to order 400 (800). The sensitivities are greatly enhanced at the FCC-hh, which could probe  $\delta y_d \,(\delta y_u)$ up to order 30 (70). We have then considered an off-shell Higgs probe of the charm Yukawa. In this case, we focus on the channel $VVcj$. First estimates of the HL-LHC and FCC-hh sensitivities, presented in Fig. \ref{fig:Yc}, indicate encouraging results and offer a useful starting point for more refined analyses at the LHC. Our first naive estimates indicate sensitivities on $\delta y_c$ in the diboson channel which could be realistically below order 1 at the HL-LHC, thus competitive and complementary to other charm Yukawa probes, and of the order of 20$\%$ at the FCC-hh. Interestingly, there is also the possibility to test the sign of the shift in the charm Yukawa.

\end{section}

\begin{section}*{Acknowledgements  }
The author thanks A.~Falkowski, S.~Ganguly, P.~Gras, J.~M.~No, K.~Tobioka, T.~You, M.~Son and E.~Venturini for discussions and collaboration in the related studies \cite{Falkowski:2020znk, Brooijmans:2020yij}, and in particular K.~Tobioka for enligthing discussions on the possibility to test the charm Yukawa in the $VVcj$ channel. 
\end{section}


 \end{document}